\documentclass[oneside,12pt,reqno]{amsart}
\usepackage{amssymb, amscd, amsmath}
\usepackage[all]{xy}
\usepackage{graphics}
\usepackage{textcomp}

 \theoremstyle{definition}
 
 \theoremstyle{remark}
 
\newlength{\vscaling} \newlength{\hscaling}
\usepackage{latexsym}
\usepackage{amssymb}

\newcommand{\eq}{\begin{equation}}
\newcommand{\en}{\end{equation}}

\def\NN{{\mathcal N}}

\def\P2{{P^{[2]}}}

\def\11{{\mbox{\boldmath $1$}}}

\def\eq{\begin{equation}}
\def\en{\end{equation}}

\def\eq#1\en{\begin{equation}#1\end{equation}}
\def\eqa#1\ena{\begin{eqnarray}#1\end{eqnarray}}

\begin{document}
\begin{titlepage}
\vskip 2cm

\vfill
{\Large\bf
\begin{center}
Large N and Bethe ansatz
\end{center}
}
\begin{center}
\vfill
{{\bf
 Branislav Jur\v co 
}}
\vskip 0.5 cm
MPI f\"ur Physik, Werner-Heiseberg-Institut\\
F\"oringer Ring 6, 80805 M\"unchen, Germany\\
\vskip 0.3cm
and 
\vskip 0.3cm
Theoretische Physik, Universit\"at M\"unchen\\ 
Theresienstr.\ 37,
80333 M\"unchen, Germany
\end{center}
\vfill
{\centerline{\bf Abstract}}
\vskip 0.3cm
\noindent
We describe an integrable model, related to the Gaudin magnet, and its
relation to the matrix model of Br\'ezin, Itzykson, Parisi and Zuber. Relation
is based on Bethe ansatz for the integrable model and its interpretation using
orthogonal polynomials and saddle point approximation. Large $N$ limit of the
matrix model corresponds to the thermodynamic limit of the integrable system. In
this limit (functional) Bethe ansatz is the same as the generating function for correlators of the
matrix models. 

\vfill
\hrule
\vskip 5pt
\noindent
{\footnotesize\it e-mail:
\parbox[t]{.8\textwidth}{jurco@theorie.physik.uni-muenchen.de}}
\end{titlepage}\vskip.2cm

\newpage

\setcounter{page}{1}
\section{\bf Introduction}
Matrix models in the large $N$-limit have been studied from different points of
view since the seminal papers of t'Hooft \cite{tH} and Br\'ezin, Itzykson,
Parisi and Zuber \cite{BIPZ}. There is no doubt of their importance
in 2D gravity and string theory (see e.g. \cite{zj},
\cite{difr}, \cite{Morozov} and \cite{Mald} for
reviews on their different aspects). 
Importance of CFT and of integrable systems for matrix models have also been 
known for some time (see e.g. \cite{difr}, \cite{Morozov}). 
Most recently there is a great interest in integrable spin chains and Bethe ansatz
in relation with large $N$-limit of $\NN=4$ SYM (see e.g. \cite{as}, \cite{bs},
\cite{mz} and references therein). This possibly 
may help to understand more
precisely the large $N$ conjecture of Maldacena \cite{maldcon}.

The purpose of this paper is to make precise the
relation between the large $N$-limit of the one-matrix model and Bethe
ansatz for some integrable model related to the classical rational
$r$-matrix (Gaudin type model). This could provide a toy model of the above mentioned
relation between the Heisenberg model and $\NN=4$ SYM.

We start with a very short review of the one-matrix model following \cite{zj}. Two
basic methods of its solution, the saddle point approximation and the method of
orthogonal polynomials, contain implicitly all information about some particular 
integrable model. Then we recall the version of Bethe ansatz appropriate for
models related to classical $r$-matrices \cite{g}, \cite{sk}, \cite{j}. 
This can be viewed as some specific limit \cite{g} of the quantum $R$-matrix 
Bethe ansatz \cite{baxter}, \cite{g}, \cite{skk}, \cite{faddeev} which is
appropriate for Heisenberg chain type of models.
Finally we will go on with an explicit description
of the relevant integrable model and discuss algebraic as well as functional Bethe
ans\"atze
for it. We will relate in a precise way the large $N$-limit of the matrix model
and the thermodynamic limit of our integrable system. In this limit Bethe
eigenfunction give the generating function for correlators of the matrix model.
Also the norm of the Bethe eigenstate is nicely related to the large $N$
expansion the partition function $Z_N$.

\section{\bf Matrix model of BIPZ}
One-matrix model is treated in detail in many places \cite{zj}, \cite{difr}. So
we can be very brief. The task is to perform the integral over $N\times N$
hermitian matrix $M$
\eq
Z_N= \int d^{N^2}\,M {\mbox{exp}}\,[-(N/g){\mbox{Tr}}\,V(M)]\,,\label{Z}
\en
where $V(M)$ is a general polynomial potential $V(M) = M^2 +
\sum_{k\geq 3}\beta_k\lambda^k$ and the integration is over $N^2$ independent
real variables which are the real and imaginary parts of matrix elements of $M$.
The standard strategy is to diagonalize the matrix $M$, i.e. write
$M=U{\mbox{diag}}(\lambda_1,...,\lambda_N)U^{\dagger}$ with an unitary $U$.
Noticing that the integrand depends only on $N$ eigenvalues $\lambda_i$ and
integrating out the unitary group one gets
\eq
Z_N=\int \prod_{i=1}^{N} d\lambda_i\, {\mbox{exp}}\,[-(N/g)\sum_i V(\lambda_i) +
\sum_{i\neq j} {\mbox{ln}}(\lambda_i -\lambda_j)]\,.
\en
The logarithmic part in the exponent justifies the
saddle point approximation for $N\rightarrow \infty$. This gives the
configuration
with the minimal energy
\eq \label{minimum}
\frac{2g}{N}\sum_{j\neq i}\frac{1}{\lambda_i -\lambda_j} = V'(\lambda_i)\,.
\en
In order to solve equation (\ref{minimum}), it is convenient to introduce the
resolvent function
\eq
W(z)=\frac{1}{N}\sum_i\left\langle\frac{1}{z-\lambda_i} \right\rangle\,.
\en
One also introduces the averaged density of eigenvalues $\rho(\lambda)$
\eq
\rho(\lambda)=\frac{1}{N}\sum_i\left\langle 
\delta(\lambda-\lambda_i)\right\rangle\,.
\en
If we assume for simplicity that the potential $V$ has only one well (or at
least one deepest well), then the support of $\rho(\lambda)$ will be in some 
interval $[a,b]$ (in general it can be a collection of disjoint intervals).
We have relations
\eq
W(z)=\int_a^bd\lambda \frac{\rho(\lambda)}{z-\lambda}
\en 
and
\eq \label{density}
\rho(\lambda) = -\frac{1}{2\pi i}(W(\lambda+i0) - W(\lambda-i0))\,.
\en
The solution for $W(z)$ is
\eq
W(z)=\frac{1}{4\pi i}\oint dx \frac{V'(x)}{z-x}\frac{\sqrt{(z-a)(z-b)}}
{\sqrt{(x-a)(x-b)}}\,,
\en 
where the integration contour encloses the cut $[a,b]$.
Finally the solution to 
(\ref{minimum}) is given by the density (\ref{density}).
See the above cited references and \cite{eynard} for more details concerning the solution of 
this equation.

An alternative method of solving (\ref{Z}) is to use orthogonal polynomials
$P_n(\lambda)=\lambda^n +...$ with respect to the
measure
\eq
{\mbox{exp}}\,[-(N/g)V(\lambda)] d\lambda\,. \label{measure}
\en
Partition function $Z_N$ is then simply expressed in terms of their norms squared
$s_n=\langle P_n|P_n\rangle$ as
\eq
Z_N=N!\,s_0...s_{N-1}. \label{sss}
\en
This follows easily from the integral representation for $P_n$ which will be given
later (\ref{ort1}), (\ref{ort2}). 
See again \cite{zj}, \cite{difr}, \cite{eynard} for a more explicit method of determination of these
polynomials. This is based on recurrence relations between orthogonal polynomials
which can be solved in the large $N$-limit. 
This little knowledge about the one-matrix model is enough to understand it as
an quantum integrable model when $N$ is large.
\section{\bf Bethe ansatz}
Before describing our integrable model we give a brief review of the relevant
Bethe ansatz following \cite{j}. Let $L(\lambda)$ be a $2\times 2$ matrix
\eq 
\left( \begin{matrix}
A(\lambda) & \,\,\,B(\lambda)\\
C(\lambda) & -A(\lambda)
\end{matrix}\right)
\en
its matrix elements being operators depending on complex parameter $\lambda$,
acting in some Hilbert space ${\mathcal{H}}$. We assume the following form of
the commutation relations for the matrix elements of $L(\lambda)$
\eq
[L(\lambda)\otimes I, I\otimes L(\mu)] + 
[r(\lambda -\mu), L(\lambda)\otimes I + I\otimes L(\mu)]=0\,,\label{com}
\en
where $I$ is the unit $2\times 2$ matrix and $r(\lambda-\mu)$ is the classical 
$r$-matrix
\eq 
\frac{1}{\lambda} \left( \begin{matrix}
1 & 0 & 0 & 0\\
0 & 0 & 1 & 0\\
0 & 1 & 0 & 0\\
0 & 0 & 0 & 1
\end{matrix}\right)\,.
\en
Note that the commutator in (\ref{com}) is a matrix commutator respecting the
operator nature of matrix elements of $L(\lambda)$. Further, we put
\eq
T(\lambda) = \frac{1}{2} {\mbox{Tr}}\, L(\lambda)^2\,,
\en
where Tr is the ``$2\times 2$ matrix trace''. as a consequence of (\ref{com})
we have the following useful commutation relations
\begin{eqnarray}
&\,&[A(\lambda), B(\mu)]  = -\frac{1}{\lambda -\mu}B(\mu) + \frac{1}{\lambda
-\mu}B(\lambda)\label{3}\,,\\
&\,&[B(\lambda), B(\mu)] = 0\,,\\
&\,&[B(\lambda), C(\lambda)] = 2A'(\lambda)
\end{eqnarray}
and
\begin{eqnarray}\label{tl}
[T(\lambda), B(\mu)]&=& \frac{1}{\lambda -\mu}A(\mu)B(\lambda) - 
\frac{1}{\lambda -\mu}B(\mu)A(\lambda)\nonumber\\ & - &
\frac{1}{\lambda -\mu}A(\lambda)B(\mu) + \frac{1}{\lambda
-\mu}B(\lambda)A(\mu)\,.
\end{eqnarray}
Operator $T(\lambda)$ is the generating function of integrals of motion in
involution as 
\eq
[T(\lambda), T(\mu)]=0\,.
\en
Next we assume that there is the so-called pseudo-vacuum $|0\rangle \in {\mathcal
H}$ such that
\begin{eqnarray}
&\,& C(\lambda)|0\rangle = 0 \,,\\
&\,& A(\lambda)|0\rangle = a(\lambda)|0\rangle 
\end{eqnarray}
and hence
\eq
T(\lambda)|0\rangle = \left(a(\lambda)^2 + a'(\lambda)\right) |0\rangle \,
\en
holds true. Algebraic Bethe ansatz (ABA) is then given by a family of vectors
\eq
|\lambda_1,...,\lambda_N\rangle=B(\lambda_1)...B(\lambda_N)|0\rangle
\,.\label{ABA}
\en
These are, as easy verified, with help of (\ref{com}--\ref{tl}), eigenstates of
$T(\lambda)$ with eigenvalues $\tau(\lambda)$
\eq
\tau(\lambda)= - 2\sum_i \frac{a(\lambda)}{\lambda -\lambda_i} +\sum_{i\neq j}
\frac{1}{\lambda -\lambda_i}\frac{1}{\lambda -\lambda_j}+
\left(a(\lambda)^2 + a'(\lambda)\right)  \label{tau}
\en
if and only if the numbers $\{\lambda_1,...,\lambda_N\}$ satisfy Bethe 
conditions
\eq
\frac{1}{2}\varphi_i \equiv a(\lambda_i) - \sum_{j\neq i}
\frac{1}{\lambda_i -\lambda_j}=0 \,. \label{bcon}
\en
The norm squared is equal to
\eq \label{normsq}
\langle \lambda_1,...,\lambda_N | \lambda_1,...,\lambda_N\rangle =
{\mbox{det}}\,\left|\frac{\partial \varphi_i}{\partial \lambda_j }\right|\,.
\en
If we introduce the polynomial $q(\lambda) = (\lambda -\lambda_1)...(\lambda
-\lambda_N)$ equation (\ref{bcon}) can be rewritten as differential equation
\eq
q'' - 2aq' + (a^2 -a')q = \tau q\,.\label{Q}
\en
\section{\bf The model and its relation to BIPZ at large N}
Comparing Bethe conditions (\ref{bcon}) and the minimal energy condition
(\ref{minimum}) we see the first
hint that there might be an integrable model hidden in the large $N$-limit of
the matrix model. To describe this integrable system we need to give a
representation of the algebra of (\ref{com}). Let us fix an arbitrary integer
$K$ and consider the space of functions of $K$ variables
$\{x_{-1},...,x_{-K}\}$. Our operators $A(\lambda)$, $B(\lambda)$ and
$C(\lambda)$ will be polynomials of orders $K$, $K-1$ and $K-1$, respectively,
acting on this space
\eq
A(\lambda) = \sum_{n={-1}}^{-K-1}A_n\lambda^{-n-1}\,, \hskip 0.5cm
B(\lambda) = \sum_{n={-1}}^{-K}B_n\lambda^{-n-1}\,, \hskip 0.5cm
C(\lambda) = \sum_{n={-1}}^{-K}C_n\lambda^{-n-1}\,,
\en
with coefficients
\begin{eqnarray}
&\,& A_n = -x_m \partial_{m-n} + a_n \,,\\
&\,& B_n = x_n\,,\\
&\,& C_n = -x_m\partial_{m-n-l}\partial_l + 2a_m\partial_{m-n}\,.
\end{eqnarray}
Here the summation rule is assumed and sums run over all integers from $-\infty$
to $K$, assuming $\partial_n = 0$ and $x_n=0$ whenever $n >-1$ and $\{a_{-1},..
a_{-K-1}\}$  are arbitrary real constants. Note that $A_{-K-1}=a_{-K-1}$: we
will assume this to be nonzero. What we have here is actually the free field
representation of a factor algebra of $\widehat{sl(2)}$ with $c=0$. ABA is
applied directly to our model. Obviously pseudo-vacuum is the constant function
$|0\rangle =1$ and we have
\eq
a(\lambda) = \sum_{n={-1}}^{-K-1}a_n\lambda^{-n-1}\,.
\en

To our integrable model also Sklyanin's functional Bethe ansatz (FBA)
is applicable. FBA in its full generality doesn't need a pseudo-vacuum. In our
case it gives nicely the separation of variables. We will not repeat here the general
discussion of \cite{sk} and just state the result.
Since $B(\lambda)$ is a polynomial of order $K-1$,
\eq
B(\lambda) = x_{-K}(\lambda -u_1)...(\lambda -u_{K-1})\,,
\en
$x_{-K}$ and its zeros $\{u_1,...,u_{K-1}\}$ are independent as functions of
variables 
$\{x_{-1},...,x_{-K}\}$. We perform the change of variables 
\eq
\{x_{-1},...,x_{-K}\}\mapsto \{u_1,...,u_{K-1},x_{-K}\}\,. 
\en
If
$q(u)=(u-\lambda_1)...(u-\lambda_N)$
is a polynomial solution of order $N$ to (\ref{Q}) then the product
\eq
\psi(u_1,...,u_{K-1})= x_{-K}^N\,q(u_1)...q(u_{K-1}) \label{psi}
\en
is an eigenfunction of $T(\lambda)$ and the corresponding eigenvalue is given by
(\ref{tau}).
The relation between the two version of Bethe ansatz is pretty obvious in our
case. We go form (\ref{ABA}) to (\ref{psi}) as follows
\begin{eqnarray}
&&| \lambda_1,...,\lambda_N\rangle = B(\lambda_1)...B(\lambda_N).1=
x_{-K}^N\prod_{i=1}^N\prod_{j=1}^{K-1}(\lambda_i-u_j)\nonumber\\&&=x_{-K}^N
\prod_{j=1}^{K-1}\prod_{i=1}^N(\lambda_i-u_j)=
(-1)^{N(K-1)} x_{-K}^N\prod_{j=1}^{K-1}q(u_j) \nonumber\\&&=
(-1)^{N(K-1)}x_{-K}^N\psi(u_1,...,u_{K-1})\,.
\end{eqnarray}
I both cases (algebraic or functional) the Bethe condition (\ref{bcon}) just
makes sure that poles of $\tau(\lambda)$ in points $\{\lambda_1,...,\lambda_N
\}$ vanish. This must happen as the operator $T(\lambda)$ doesn't have poles in
these points. If we put $a_{-1}=0$, $a_{-2}=N/g$ and
$a_{-n}=\frac{nN}{2g}\beta_n$
for $n\geq 3$, so that we have $a=\frac{N}{2g}V'$, then Bethe conditions 
(\ref{bcon}) and the minimum energy condition (\ref{minimum}) are identical.

We can go even further. Following Szeg{\"o}'s book on orthogonal polynomials
(and being inspired by matrix model) we consider a system of orthonormal
polynomials $p_N(u)$ with respect to the measure (\ref{measure}) given as
\begin{eqnarray}\label{ort1}
&&p_n(u)=\frac{(D_{n-1}D_n)^{-1/2}}{n!}\times \\
&&\int \prod_{i=1}^{n}
d\alpha_i\,(u-\alpha_1)...(u-\alpha_n)\,{\mbox{exp}}\,[-\sum_i(N/g)V(\alpha_i) +
\sum_{i\neq j}{\mbox{ln}}(\alpha_i -\alpha_j)]\,,\nonumber
\end{eqnarray}
where
\eq \label{ort2}
D_n=\frac{1}{(n+1)!}\int \prod_{i=1}^{n+1}
d\alpha_i\,{\mbox{exp}}\,[-\sum_i(N/g)V(\alpha_i) +
\sum_{i\neq j}{\mbox{ln}}(\alpha_i -\alpha_j)].
\en
Obviously we do have 
\eq 
P_n(u)=
\left[\frac{D_{n}}{D_{n-1}}\right]^{1/2}p_n(u)
\en
and (\ref{sss}) follows
trivially.
Let us now assume $N\rightarrow \infty$. Again the logarithmic term in the
potential allows to apply the saddle point method to the integrals defining
$P_N(u)$, which are generating functions of correlators for matrix model.
This is done straightforwardly with the result
\eq
P_N(u) = (u-\lambda_1)...(u-\lambda_N)+O([N/g])^{-1}\,.
\en
Here $\{\lambda_1,...,\lambda_N\}$ must again minimize the energy, hence
they satisfy (\ref{minimum}).
Recalling the functional Bethe ansatz ($\ref{psi}$) we see that the polynomials
\eq
P_N(u_1)...P_N(u_{K-1})\rightarrow q(u_1)...q(u_{K-1})
\en
in the large $N$-limit and apart from an inessential factor $x^N_{-K}$ they give Bethe
eigenvectors
$\psi(u_1,..., u_{K-1})$ of $T(\lambda)$. 
Also (see (\ref{normsq}))
\begin{eqnarray}
Z_N&=&\left[\frac{2\pi}{N/g}\right]^{N/2}\langle \lambda_1...\lambda_N|\lambda_1...\lambda_N
\rangle^{-1/2} \times \nonumber \\&\,&{\mbox{exp}}\,[-(N/g)\sum_i V(\lambda_i) +
\sum_{i\neq j} {\mbox{ln}}(\lambda_i -\lambda_j)](1 + O(g/N)).
\end{eqnarray}

Although Bethe states (\ref{ABA}) or 
(\ref{psi}) are eigenstates of our integrable model for {\it{any}} $N$ here
we recovered them only in the large $N$-limit. For our representation labels
$\{a_{-2},...,a_{-K-1}\}$ this means that they grow proportionally to $N$ as this
goes to infinity. At the same time $N$ is the number of vacuum excitations. Hence we can interpret the large $N$-limit as the thermodynamic limit of
our integrable system. Clearly in this limit the polynomial $q(u)$ is the
{\it{generating function}} of correlators for the matrix model.

\end{document}